\documentclass[aps,prb,twocolumn]{revtex4}
\usepackage{epsf}

\begin{document}

\title{
Kondo screening in $d$-wave superconductors in a Zeeman field \\
and implications for STM spectra of Zn-doped cuprates
}

\author{Matthias Vojta, Robert Zitzler, Ralf Bulla, and Thomas Pruschke}
\affiliation{
Theoretische Physik III, Elektronische Korrelationen und
Magnetismus, Universit\"at Augsburg, 86135 Augsburg, Germany
}
\date{February 19, 2002}

\begin{abstract}

We consider the screening of an impurity moment
in a $d$-wave superconductor under the influence of
a Zeeman magnetic field.
Using the Numerical Renormalization Group technique, we
investigate the resulting pseudogap Kondo problem, in
particular the field-induced crossover behavior in the
vicinity of the zero-field boundary quantum phase
transition.
The impurity spectral function and the resulting changes
in the local host density of states are calculated,
giving specific predictions for high-field scanning tunneling microscopy
measurements on impurity-doped cuprates.

\end{abstract}

\maketitle


\section{Introduction}

Impurity effects provide valuable information
about bulk properties of correlated electronic systems
such as cuprate superconductors.
Remarkably, non-mag\-ne\-tic impurities like Zn and Li can induce a
local moment on the neighboring lattice sites at intermediate
energies~\cite{alloul,sisson,bobroff,julien,entity}.
On the other hand, Zn is known to act as pair breaker and strong
scatterer, e.g. in low-tempera\-ture transport measurements, and at
present it is not clear whether this behavior arises from
bare potential scattering only, or if other effects like Kondo physics
play a role.

Local impurity moments -- arising either from magnetic impurities or
induced by non-magnetic defects -- interact with the low-energy
bulk excitations.
For cuprates at temperatures $T$ below the superconducting $T_c$,
these are the fermionic quasiparticles
of the $d$-wave superconductor and collective bosonic
spin fluctuations.
The coupling to the spin fluctuations
mainly plays a role at
intermediate energy scales of the order of the spin gap
(an estimate is given by the position of the so-called
resonance mode at 40 meV) --
their effect on the impurity dynamics has been discussed
recently in Ref.~\onlinecite{impmag}.

For smaller energies, the dynamics of the impurity moments
is governed by the interaction with the fermionic Bogoliubov
quasiparticles.
Importantly, their single-particle density of states (DOS) vanishes
at the Fermi energy, therefore the physics is quite
distinct from the usual Kondo effect in metals~\cite{hewson}.
Extensive studies of the pseudogap Kondo
model~\cite{withoff,GBI,tolya,MVRB,OSPG},
where the host DOS follows a power-law near the Fermi level,
$\rho(\omega) \sim |\omega|^r$ ($r=1$ for $d$-wave superconductors),
have shown the existence of a boundary quantum phase transition
between a free moment and a screened moment as the Kondo
coupling is increased.
Applied to the cuprates, we have argued~\cite{MVRB} that this can
lead to a strongly doping-dependent Kondo temperature, caused by the
doping dependence of the size of the $d$-wave gap.
These theoretical findings are consistent with NMR experiments
\cite{bobroff}
measuring the local susceptibility near a non-magnetic impurity.
This susceptibility has been found to show Curie-Weiss-like behavior,
with a Weiss temperature being finite at optimal doping but
apparently vanishing for underdoped samples --
this implies the existence of unscreened moments in the underdoped
regime down to temperatures of order 1 K.

Interesting impurity signatures have also been observed in
recent scanning tunneling microscopy (STM) experiments
of the high temperature superconductor ${\rm
Bi}_2 {\rm Sr}_2 {\rm CaCu}_2 {\rm O}_{8+ \delta}$ (BSCCO)
\cite{seamuszn,seamusni}.
Studies in the vicinity of Zn impurities on the Cu sites~\cite{seamuszn}
have found a large peak in the
differential conductance at small energies of 1--2 meV
near the Zn site, with a characteristic spatial variation of
the signal.
This peak could represent a quasi-bound state in a potential
scattering model\cite{bala,tsu,alan},
but such a state appears at low energies only for a
range of large potential values
depending upon microscopic details \cite{alan}.
Recently, it has been proposed \cite{tolya,MVRB} that some of the
properties of the Zn resonance can be explained by the Kondo spin
dynamics of the magnetic moment induced by Zn.
The screening of this moment by the Bogoliubov quasiparticles
provides a natural low-energy scale (the Kondo temperature)
explaining the energetic location of the peak seen in STM --
in this case the potential scattering model might still apply
as an {\em effective} model at low temperatures.
However, not all features of the Zn resonance can be
understood within the existing models, and important ingredients
in the theoretical description, e.g., the actual tunneling
path of the electrons from the tip to the CuO$_2$ plane \cite{filter},
are not well known, calling for further experimental clarification.

In this paper we study the screening of an impurity moment
embedded in a $d$-wave superconductor in a Zeeman magnetic
field;
we do neglect the changes in the quasiparticle spectrum
due to the superflow induced by vortices which is appropriate
for in-plane magnetic fields (parallel to the copper oxide layers)
and for impurities located in between two vortices.
In particular, we calculate the field-dependent impurity spectral
density -- based on these results, we propose that STM tunneling
experiments in high magnetic fields will be able to distinguish between
the possible sources of the observed impurity STM signal, and
thus contribute to clarify the nature of impurity scattering in
cuprate superconductors.

\section{Model}

We begin by describing our model Hamiltonian, $H=H_{\rm
BCS}+H_{\rm imp}$, for a superconductor with a single impurity.
The first term describes the host superconductor, which we model by a
simple BCS Hamiltonian
\begin{equation}
H_{\rm BCS} = \sum_{\bf k} \Psi^{\dagger}_{\bf k} \left[
(\varepsilon_{\bf k} - \mu) \tau^z + h_{\rm bulk} + \Delta_{\bf k} \tau^x \right]
\Psi_{\bf k}.
\end{equation}
Here $\Psi_{\bf k} = (c_{{\bf k}\uparrow}, c_{-{\bf k}\downarrow}^{\dagger})$ is
a Nambu spinor at momentum ${\bf k}=(k_x, k_y)$ in standard notation.
$\tau^{x,y,z}$ are Pauli matrices in particle-hole space, and $\mu$ is
the chemical potential. For the kinetic energy, $\varepsilon_{\bf k}$, we
have first ($t$), second ($t^{\prime}$), and third
($t^{\prime\prime})$ neighbor hopping,
while we assume a
$d$-wave form for the BCS gap function $\Delta_{\bf k} =
(\Delta_0/2)
(\cos k_x - \cos k_y )$.
The spin quantization axis lies in the direction of
the (in-plane) Zeeman field $h_{\rm bulk}$.

The Green's function of the
conduction electrons described by $H_{\rm BCS}$ is
$
G^0 ({\bf k}, \omega) =
[\omega - (\varepsilon_k -\mu) \tau^z - h_{\rm bulk} - \Delta_k \tau^x ]^{-1}
$
in matrix notation.
The influence of a Zeeman magnetic field on a $d$-wave BCS superconductor
has been studied in Ref.~\onlinecite{yang}.
For low temperatures and small fields, $h_{\rm bulk} \ll \Delta_0$,
the field-induced change in the gap function $\Delta_{\bf k}$ can
be neglected.
Thus the main effect of the Zeeman field is to split the
quasiparticle spectrum by $\pm h_{\rm bulk}$, which leads to a finite
DOS at the Fermi level.

We will not consider the coupling of the external field to the
orbital motion of the electrons -- the primary effect of the field-induced
superflow would be an additional shift in the quasiparticle spectrum.
Neglecting this Doppler effect is approximately justified for
fields applied parallel to the CuO$_2$ planes of the
system:
Due to the layered crystal structure and strong anisotropy,
the orbital effect of the magnetic field is much weaker for
in-plane fields, and vortex effects are likely to be small~\cite{yang}.
Furthermore, we expect our approximation to apply also
for fields in $c$-axis direction, i.e., perpendicular to the
planes, for impurities located in between two vortices
where the superflow tends to cancel and hence the Doppler shift
is small.

\begin{figure}[t]
\epsfxsize=3.5in
\centerline{\epsffile{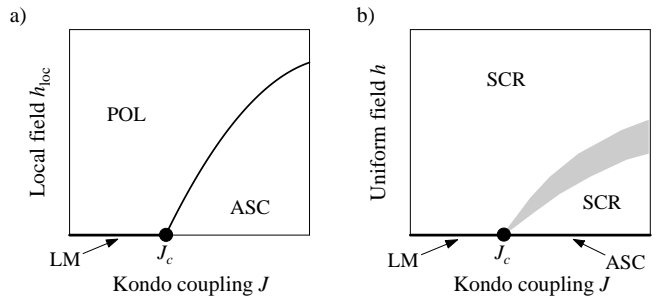}}
\caption{
Schematic zero-temperature phase diagrams for the asymmetric pseudogap
Kondo model in a Zeeman field, deduced from NRG.
a) Local field $h_{\rm loc}$ only, i.e., for $h_{\rm bulk}\!=\!0$.
There is a line of continuous phase transitions at finite fields,
terminating in the multicritical point at $J\!=\!J_c$, $h_{\rm loc}\!=\!0$.
(The phases LM, ASC, and POL are explained in the text.)
b) Uniform field $h\!=\!h_{\rm loc}\!=\!h_{\rm bulk}$.
Here, the asymptotic low-$T$, low-frequency behavior at finite fields
is always that of a conventional Kondo-screened spin in a magnetic
field (SCR phase), but there is a sharp crossover separating regions which
are dominated by ASC and POL behavior at finite energies.
}
\label{figphd1}
\end{figure}

The impurity, located at ${\bf r}_0 = (0,0)$,
is described by an effective model consisting of
a scattering potential $U$ and a Kondo term for
the spin-1/2 impurity spin $\bf S$:
\begin{equation}
H_{\rm imp} =
\sum_{\bf R \in \cal{N}} J_{\bf R} {\bf S} \cdot {\bf s}_{\bf R}
+ S^z h_{\rm loc}
+ U  c^{\dagger}_{\alpha} ({\bf r}_0) c_{\alpha} ({\bf r}_0).
\label{Himp}
\end{equation}
where ${\bf s}_{\bf R} = N_s^{-1} \sum_{\bf kk'\alpha\beta}
{\rm e}^{{\rm i}({\bf k}-{\bf k'}){\bf R}} c^\dagger_{{\bf k}\alpha} \frac{1}{2} {\bf \sigma}_{\alpha\beta}
c_{{\bf k '}\beta}$ is the conduction band spin operator at site ${\bf R}$.
As discussed in Refs.~\onlinecite{impmag,tolya,MVRB} we will consider
two cases:
(i) a local magnetic moment where ${\cal N} = \{{\bf r}_0\}$, and
(ii) a spatially extended magnetic moment (modelling Zn) where
${\cal N}$ is a set of sites in the neighborhood
of ${\bf r}_0$; guided by NMR experiments~\cite{julien}
we take ${\cal N}$ to be the four neighbors of
${\bf r}_0$, i.e., $J_{\bf R} = J$ for
${\bf R} - {\bf r}_0 = (\pm 1, 0), (0,\pm 1)$ and $J_{\bf R}=0$ otherwise.
Note that this four-site model describes
an induced moment fluctuating as a {\em single} entity, in
agreement with spin relaxation experiments \cite{entity}.
The Kondo term in the four-site case defines a multichannel
Kondo problem; however, the
low-energy physics in the present situation is dominated by a
single channel given by the $d$-wave-like linear combination of
the conduction electrons on the four neighboring sites~\cite{tolya,MVRB}.

The potential scattering term in (\ref{Himp})
can be accounted for exactly:
\begin{eqnarray}
G({\bf r},{\bf r}^{\prime},\omega) =
G^0 ({\bf r}-{\bf r}^{\prime},\omega)
+ U G^0 ({\bf r}-{\bf r}_0,\omega) \nonumber \\
 \times \, \tau^z [ 1 -
U G^0 ((0,0),\omega) \tau^z ]^{-1} G^0 ({\bf r}_0-{\bf
r}^{\prime},\omega).
\label{scatgf}
\end{eqnarray}
The Kondo problem we consider in the following is then
defined by the spin-dependent DOS seen
by the impurity.
For a single-site impurity we have
\begin{equation}
\rho_{{\rm eff},\alpha} (\omega) = -{\rm Im}
G_{\alpha\alpha}({\bf r}_0, {\bf r}_0, {\rm sgn}(\alpha) \omega) \,/\, \pi
\label{rho1site}
\end{equation}
with $\rho_{{\rm eff},\alpha} (\omega) \sim |\omega|$ at zero field,
whereas the extended four-site impurity leads to
\begin{equation}
\rho_{{\rm eff},\alpha} (\omega) = -\frac{1}{4\pi}
{\rm Im}
\sum_{\bf R, R'} \varphi_{\bf R} \varphi_{\bf R'}
G_{\alpha\alpha}({\bf R}, {\bf R'}, {\rm sgn}(\alpha) \omega)
\label{rho4site}
\end{equation}
which behaves as $\rho_{{\rm eff},\alpha} (\omega) \sim |\omega|^3$ at
$h_{\rm bulk}=0$.
Here, $\varphi_{\bf R} = +[-]1$ for ${\bf R} - {\bf r}_0 = (\pm 1, 0) [(0,\pm 1)]$
describe the $d$-wave channel of the four-site impurity,
$G$ is the $\Psi$ Green's function (\ref{scatgf}) in Nambu notation,
and ${\rm sgn}(\alpha) = \pm 1$ for $\alpha = \uparrow,\downarrow$.


\section{Boundary critical points and crossover behavior}

In the following, we describe our numerical results for
the above Kondo model in an external field,
obtained with Wilson's Numerical Renormalization Group (NRG)
technique~\cite{nrg}.
The zero-field pseudogap Kondo problem in the presence of
particle-hole asymmetry
is known to show a boundary quantum phase transition between
an asymmetric strong-coupling (ASC) phase with a screened impurity moment
and a local moment (LM) phase with Curie behavior of the impurity
susceptibility~\cite{GBI,MVRB}.

\begin{figure}[t]
\epsfxsize=3.5in
\centerline{\epsffile{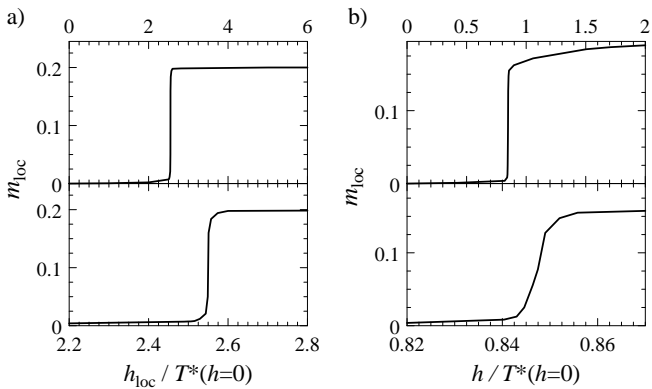}}
\caption{
NRG results for the local impurity magnetization
$m_{\rm loc} = \langle S^z \rangle$, calculated for a four-site impurity in
a cuprate band structure
($t=-0.15$ eV, $t'=-t/4$, $t''=t/12$, $\Delta_0=0.04$ eV, $\mu=-0.14$ eV)
with $J_c \approx 0.23$ eV for $U=0$ ~\protect\cite{jc},
as a function of the applied field.
a) Local field $h_{\rm loc}$ only, for a $J>J_c$ with $T^\ast \approx 5$ K.
At the phase transition 
the magnetization follows a power law,
$|m_{\rm loc}-m_{{\rm loc},c}| \sim |h_{\rm loc}-h_{{\rm loc},c}|^\zeta$,
with $\zeta\approx 0.1$.
b) Uniform field $h\!=\!h_{\rm loc}\!=\!h_{\rm bulk}$, for a $J>J_c$ with $T^\ast \approx 25$ K.
Here, the magnetization evolves smoothly through the crossover region of
Fig.~\protect\ref{figphd1}b.
The lower figures show zooms of the upper ones.
}
\label{figmag1}
\end{figure}

We start by considering the case of a local field only, $h_{\rm loc} \neq 0$
but $h_{\rm bulk}=0$, such that the host DOS is unaffected by the
field.
From the analysis of the fixed points of the NRG calculation
we have found that even in a finite local field there is still a
phase transition upon varying the Kondo coupling.
The $T=0$ phase diagram deduced from NRG is schematically shown in Fig.~\ref{figphd1}a --
there is a line of finite-field transitions ending in the
zero-field transition point known from earlier work
\cite{GBI,MVRB} which can be viewed as a
multicritical point.
The three stable phases are easily understood:
The ASC phase at $J>J_c$ extends to
finite field (with weak spin polarization for $h_{\rm loc} \neq 0$),
whereas the LM phase at $J<J_c$ with a free impurity spin
exists at zero field only.
Applying a field at $J<J_c$ immediately leads to a phase with
strong polarization of the impurity spin, therefore we term
this phase POL.

All transitions found are continuous phase transitions,
characterized by a small energy scale
$T^\ast$, which vanishes at the transition, and defines the crossover
energy above which quantum-critical behavior is observed.
In the ASC phase at zero field, $T^\ast(J,h\!=\!0)$ can be identified with the
Kondo temperature~\cite{MVRB}, i.e., with the binding energy of the Kondo
singlet.
The transition from ASC to POL is characterized by breaking this singlet
by the external field -- consequently the critical field $h_{{\rm loc},c}$
close to the multicritical point is given by
$T^\ast(J,h\!=\!0)$ times a prefactor depending on the DOS exponent $r$.
Note that this novel {\em field-induced Kondo transition} is unique to the
pseudogap model studied here -- in the metallic host case ($r\!=\!0$)
the application of a field leads only to a gradual suppression of
the Kondo effect~\cite{KONDOF}.

From the above discussion we expect that a {\em uniform} Zeeman field
$h_{\rm bulk}\!=\!h_{\rm loc}\!=\!h$ removes the critical point,
because a finite $h_{\rm bulk}$ leads to a finite DOS at the Fermi
level.
Our numerical calculations show that this is indeed the case -- for any
finite $h$ the asymptotic zero-temperature behavior is that of a conventional
(i.e. metallic-host) Kondo-screened spin in a magnetic field.
However, the characteristic energy scale below which this behavior occurs
is tiny, as it is exponentially small in the field-induced DOS at the
Fermi level.
(The same applies to effects arising from the small low-energy DOS induced
by a finite concentration of impurities, i.e., the so-called impurity band~\cite{MVRB}.)
Therefore, at practically accessible temperatures the finite-field behavior
is determined by the physics of the POL and ASC phases described above,
and both regimes are now separated by a sharp crossover, Fig.~\ref{figphd1}b.
This crossover field turns out to be somewhat smaller
than the critical {\em local} field in Fig.~\ref{figphd1}a;
their ratio depends on the field-induced DOS at the Fermi level
(i.e. the DOS exponent $r$).

The properties of the phases are reflected in the
field-induced impurity magnetization, Fig.~\ref{figmag1}.
This quantity is finite for any non-zero field, but shows a power-law
singularity at the transition in the local-field case (Fig.~\ref{figmag1}a).
Note that the behavior of $m_{\rm loc}$ near the transition is not
universal but depends on $J$ -- the reason is that the
finite-field transitions actually represent a {\em line} of critical fixed
points in the RG sense.
In the uniform-field case, Fig.~\ref{figmag1}b, the singularity in $m_{\rm loc}$
is replaced by a relatively sharp crossover.


\section{Impurity $T$ matrix and STM}

Turning to the dynamic properties of the Kondo impurity,
we now discuss the conduction electron $T$ matrix, $T(\omega)$, which is
the important input quantity for the calculation of the local
DOS measured by STM.
The $T$ matrix spectral density, $\rho_{T_\alpha}(\omega) = - {\rm Im} T_\alpha(\omega)/\pi$,
in zero external field has been studied in Ref.~\onlinecite{MVRB}
using NRG.
It was found that $\rho_T(\omega) \sim |\omega|^r$ in the stable
LM and ASC phases.
The critical point at $J_c$ is characterized by
$\rho_T(\omega) \sim |\omega|^{-r}$ (for $r\leq 1$) \cite{MVRB,OSPG}.
The crossover from quantum-critical $|\omega|^{-r}$ behavior at high energies
to $|\omega|^{r}$ behavior at low energies leads to a characteristic
peak in the spectral density on one (!) side of the Fermi level
at an energy $T^\ast(J,h\!=\!0)$, which can be viewed as Kondo peak
(note that the problem at hand is intrinsically particle-hole asymmetric).

\begin{figure}[t]
\epsfxsize=3.5in
\centerline{\epsffile{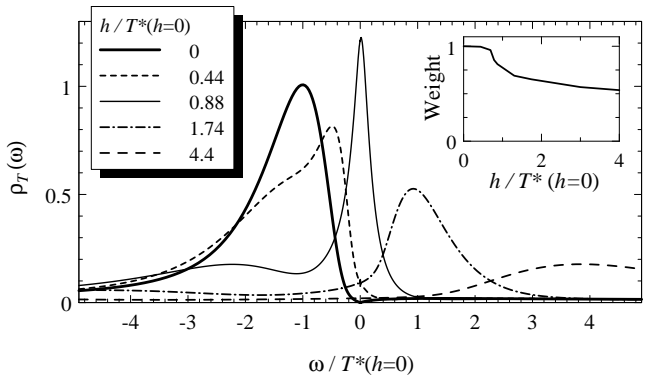}}
\caption{
Typical NRG results for the $T$ matrix spectral den\-sity,
$\sum_\alpha \rho_{T_\alpha} (\omega)$, at fixed $J\!>\!J_c$ for different
uniform fields $h$, calculated for a local impurity \cite{sigma}.
The height of the Kondo peak reaches its maximum in the
crossover region of Fig.~\protect\ref{figphd1}b and rapidly decreases
for higher fields.
Inset:
Normalized weight of the Kondo peak,
integrated from $\omega = -10 T^\ast$ to $10 T^\ast$.
}
\label{figsp1}
\end{figure}

For finite field, the spin degeneracy is lifted: applying a small field
to the system with $J > J_c$ splits the Kondo peak,
in a manner roughly similar to the field-induced peak splitting in
the conventional (metallic) Kondo effect~\cite{KONDOF}.
For $h_{\rm bulk}=0$ the low-frequency limit of $\rho_T(\omega)$
still behaves as $|\omega|^r$;
of course, finite $h_{\rm bulk}$ will lead to a finite value of
$\rho_T(\omega\!=\!0)$.
The simple peak-splitting picture is modified when the field becomes
comparable to the zero-field peak energy, i.e., close
to the transition/crossover lines in Fig.~\ref{figphd1}.
In the case of a purely local field,
we observe that the spectral peak for one spin direction
approaches the Fermi level and becomes sharper as
$|h_{\rm loc}-h_{{\rm loc},c}|$ is decreased;
it diverges {\em at} the phase transition as $|\omega|^{-r}$, whereas
the spectrum for the opposite spin broadens and remains non-critical.
For a uniform field, this effect is again smeared out:
in the crossover region of Fig.~\protect\ref{figphd1}b the spectral
function shows a peak with a finite, small width {\em at} the Fermi
level, see Fig.~\ref{figsp1}.
Remarkably, the weight of the Kondo peak (inset of Fig.~\ref{figsp1})
stays nearly constant at small fields (ASC regime), but rapidly
decreases (POL regime) once the crossover region is passed.

Applying a field to a system with $J<J_c$ immediately suppresses
the spectral function for one spin species whereas the other one shows
a smaller peak
which moves away from the Fermi level and gradually looses intensity --
so there are no drastic effects as suggested by the phase diagram.

\begin{figure}
\epsfxsize=3.5in
\centerline{\epsffile{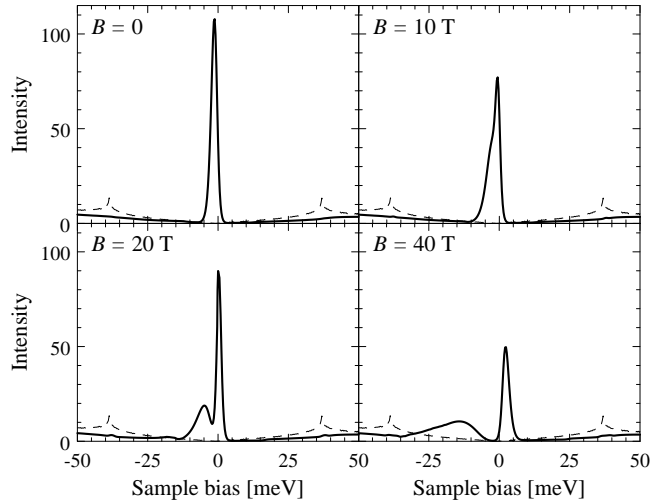}}
\caption{
Field evolution of the tunneling DOS for the four-site
Kondo impurity, measured on the impurity site, without
taking into account a possible filter effect~\protect\cite{filter}.
The host paramaters are as in Fig.~\protect\ref{figmag1},
the Kondo coupling is chosen to yield a zero-field peak at
$-1.4$ meV ($T^\ast=15$ K), the potential scattering is $U=0$.
The results have been broadened corresponding to a measurement
temperature of 4.5 K; the dashed line shows the bulk DOS.
(The external field $B$ is related to $h$ by $h=g \mu_B B / 2$.)
}
\label{figstm1}
\end{figure}

Using the $T$ matrix from the NRG method~\cite{sigma} it is straightforward to determine
the local conduction electron DOS~\cite{MVRB} as measured by
STM.
Fig.~\ref{figstm1} shows a set of calculated on-site tunneling
spectra for a four-site impurity in a cuprate band structure, which
is supposed to describe the moment induced by a Zn impurity
in BSCCO.
For a zero-field peak position of $-1.4$~meV, the crossover
region of Fig.~\ref{figphd1}b is reached at an in-plane field
of approximately 20~T. For larger fields, the peak moves to
positive bias and looses intensity similar to the inset of
Fig.~\ref{figsp1}.
The spatial intensity pattern of the signal is nearly
field-independent; it is alternating from site to site, i.e.,
the nearest neighbor of the impurity
has a very small intensity, whereas the second neighbor has again
higher intensity~\cite{tolya,MVRB,bala}.
Fig.~\ref{figstm1} shows results for zero potential scattering $U$,
we note that no qualitative modifications occur for small and moderate
$U$ values up to the hopping energy $t$, see also 
Ref.~\onlinecite{MVRB}.

\section{Conclusions}

Summarizing, we have studied the effect of a Zeeman field on
the Kondo screening of impurity moments embedded in a host
with a pseudogap density of states.
The application of a local magnetic field leads to a
novel field-induced boundary phase transition, associated
with breaking up the Kondo singlet -- this transition has
characteristic signatures in the local magnetization and
the impurity spectral function.
In the context of Zn impurities in cuprate $d$-wave superconductors
we predict a characteristic evolution of the impurity STM spectrum
with applied field which can be tested in future experiments.
The spectra in Fig.~\ref{figstm1} can be contrasted to the
results for a pure potential scattering model~\cite{grimaldi},
where the scattering peak is simply split by the Zeeman
energy, but is not modified in its total weight -- therefore
these experiments will help to clarify the general role played
by non-magnetic impurities in cuprates.


\acknowledgments

We thank S. Davis, D. Morr, A. Polkovnikov, and S. Sachdev
for valuable discussions.
This research was supported by the DFG through SFB 484.



\begin{thebibliography}{999}
\bibitem{jc}
This corrects a numerical error in our earlier computation, Ref.~\protect\onlinecite{MVRB}.
\end{thebibliography}
\end{document}